\documentclass{elsart}
\usepackage{graphicx}

\def\ifm#1{\relax\ifmmode#1\else$#1$\fi}
\def\DAF{DA\char8NE}  \def\sig{\ifm{\sigma}}
\def\f{\ifm{\phi}}    
\def\ab{\ifm{\sim}}  \def\x{\ifm{\times}}
\def\gam{\ifm{\gamma}}  
\def\pt#1,#2,{\ifm{#1\x10^{#2}}}   \def\epm{\ifm{e^+e^-}}
\renewcommand{\to}{\ensuremath{\rightarrow}}
  \def\po{\ifm{\pi^0}}  \def\et{\ifm{\eta}}
\def\dif{{\rm d\hspace{.3mm}}}

\makeatletter
\newdimen\z@ \z@=0pt 
\newskip\z@skip \z@skip=0pt plus0pt minus0pt
\def\m@th{\mathsurround=\z@}
\def\ialign{\everycr{}\tabskip\z@skip\halign} 
\def\eqalign#1{\null\,\vcenter{\openup\jot\m@th              
  \ialign{\strut\hfil$\displaystyle{##}$&$\displaystyle{{}##}$\hfil
      \crcr#1\crcr}}\,}
\makeatother

\newcommand{\ep}{\mbox{$e^+$}}
\newcommand{\el}{\mbox{$e^-$}}
\newcommand{\ks}{\mbox{$K_S$}}
\newcommand{\kl}{\mbox{$K_L$}}
\newcommand{\pip}{\mbox{$\pi^+$}}
\newcommand{\pim}{\mbox{$\pi^-$}}
\newcommand{\pio}{\mbox{$\pi^{0}$}}
\newcommand{\fo}{\ensuremath{f_0}}
\newcommand{\ppg}{\mbox{$\pi\pi\gamma$}}
\newcommand{\wpi}{\mbox{$\omega\pi$}}
\newcommand{\mpp}{\mbox{$M_{\pi\pi}$}}
\newcommand{\subfo}{\mbox{$_{f\!_{0}}$}}
\newcommand{\phifog}{\mbox{$\phi\to f\!_{0}\gamma$}}
\newcommand{\phippg}{\mbox{$\phi\to\pio\pio\gamma$}}

\newcommand{\wpn}{\mbox{$\ep\el\to\omega\pio\to\pio\pio\gamma$}}
\newcommand{\dafne}{\mbox{DA$\Phi$NE}}
\newcommand{\br}{\mbox{BR}}
\newcommand{\qq}{\mbox{$q\overline{q}$}}
\newcommand{\kk}{\mbox{$K\overline{K}$}}
\newcommand{\qqqq}{\mbox{$q\overline{q}q\overline{q}$}}
\newcommand{\lint}{\mbox{$L_{\rm int}$}}
\newcommand{\aff}[2]
  {Dipartimento di Fisica dell'Universit\`a #1 e Sezione INFN, #2, Italy.}
\newcommand{\affd}[1]
  {Dipartimento di Fisica dell'Universit\`a e Sezione INFN, #1, Italy.}
\newcommand{\AmS}{{\protect\the\textfont2
  A\kern-.1667em\lower.5ex\hbox{M}\kern-.125emS}}

\begin{document}

\begin{frontmatter}

\title{\boldmath Study of the Decay \phippg\ with the KLOE Detector}

%
\collab{The KLOE Collaboration}
\author[Na]{A.~Aloisio},
\author[Na]{F.~Ambrosino},
\author[Frascati]{A.~Antonelli},
\author[Frascati]{M.~Antonelli},
\author[Roma3]{C.~Bacci},
\author[Frascati]{G.~Bencivenni},
\author[Frascati]{S.~Bertolucci},
\author[Roma1]{C.~Bini},
\author[Frascati]{C.~Bloise},
\author[Roma1]{V.~Bocci},
% ...
\address[Bari]{\affd{Bari}}
\address[Frascati]{Laboratori Nazionali di Frascati dell'INFN, 
  Frascati, Italy.}
\address[Karlsruhe]{Institut f\"ur Experimentelle Kernphysik, 
  Universit\"at Karlsruhe, Germany.}
\address[Lecce]{\affd{Lecce}}
\address[Na]{Dipartimento di Scienze Fisiche dell'Universit\`a 
  ``Federico II'' e Sezione INFN,
  Napoli, Italy}
\address[Pisa]{\affd{Pisa}}
\address[Energ]{Dipartimento di Energetica dell'Universit\`a 
  ``La Sapienza'', Roma, Italy.}
\address[Roma1]{\aff{``La Sapienza''}{Roma}}
\address[Roma2]{\aff{``Tor Vergata''}{Roma}}
\address[Roma3]{\aff{``Roma Tre''}{Roma}}
\address[StonyBrook]{Physics Department, State University of New 
  York at Stony Brook, USA.}
\address[Trieste]{\affd{Trieste}}
\address[Virginia]{Physics Department, University of Virginia, USA.}
\address[Beijing]{Permanent address: Institute of High Energy 
  Physics, CAS,  Beijing, China.}
\address[Moscow]{Permanent address: Institute for Theoretical 
  and Experimental Physics, Moscow, Russia.}
\corauth[cor1]{Corresponding author: Simona Giovannella
INFN - LNF, Casella postale 13, 00044 Frascati (Roma), 
Italy; tel. +39-06-94032697, e-mail simona.giovannella@lnf.infn.it}
\corauth[cor2]{Corresponding author: Stefano Miscetti
INFN - LNF, Casella postale 13, 00044 Frascati (Roma), 
Italy; tel. +39-06-94032771, e-mail stefano.miscetti@lnf.infn.it}
%
\begin{abstract}
%
We have measured the branching ratio BR(\phippg) with the KLOE detector 
using a sample of \ab\pt5,7, \f\ decays. \f\ mesons are produced at
\DAF, the Frascati \f-factory. We find BR(\f\to\po\po\gam)=
\pt(1.09\pm0.03_{\rm stat}\pm0.05_{\rm syst}),-4,.
We fit the two--pion mass spectrum to models to disentangle contributions
from various sources.

\end{abstract}

\begin{keyword}
$e^+e^-$ collisions \sep $\phi$ radiative decays \sep scalar mesons
\PACS 13.65.+i \sep 14.40.-n
\end{keyword}

\end{frontmatter}


The decay \phippg\ was first observed in 1998 \cite{Obs_fo}. Only two 
experiments have measured its rate \cite{SND_fo,CMD_fo}.
The measured rate is too large if \f\to\fo(980)\gam, with $\fo\to\pio\pio$,
were the dominating contribution and \fo(980) is interpreted as a \qq\ scalar 
state \cite{CIK,BrownClose}.
Possible explanations for the \fo\ are: ordinary \qq\ meson, \qqqq\ state, 
\kk\ molecule \cite{Tornqvist,Jaffe,Molecule,CIK}. 
Similar considerations apply also to the $a_0$(980) meson.
The decay \phippg\ can clarify this situation since both the branching 
ratio and the line shape depend on the structure of the \fo. We present in 
the following a study of the decay \phippg\ performed with 
the KLOE detector~\cite{KLOE} at \DAF~\cite{DAFNE}, an 
\epm\ collider which operates at a center of mass energy 
$W$=$M_\phi\sim 1020$ MeV. 
Data were collected in the year 2000 for an integrated luminosity 
$\lint\!\sim\!16$ pb$^{-1}$, corresponding to around \pt5,7, 
$\phi$-meson decays.

The KLOE detector consists of a large cylindrical drift chamber, DC, 
surrounded by a lead-scintillating fiber electromagnetic calorimeter, EMC. 
A superconducting coil around the EMC provides a 0.52 T field. 
The drift chamber~\cite{DCH}, 4~m in diameter and 3.3~m long, has 12,582 
all-stereo tungsten sense wires and 37,746 aluminum field wires. The chamber 
shell is made of carbon fiber-epoxy composite and the gas used is a 90\% 
helium, 10\% isobutane mixture. These features maximize transparency to 
photons and reduce $\kl\to\ks$ regeneration and multiple scattering. The 
position resolutions are $\sigma_{xy}$\ab150 $\mu$m and $\sigma_z$\ab~2 mm. 
The momentum resolution is $\sigma(p_{\perp})/p_{\perp}\approx 0.4\%$. 
Vertices are reconstructed with a spatial resolution of \ab3~mm. 
The calorimeter~\cite{EMC} is
divided into a barrel and two endcaps, for a total of 88 modules, and covers 
98\% of the solid angle. The modules are read out at both ends by 
photomultipliers; the readout granularity is \ab4.4\x4.4~cm$^2$, for a total of
2440 cells. The arrival times of particles and the positions in three 
dimensions of the energy deposits are obtained from the signals collected at 
the two ends. Cells close in time and space are grouped into a calorimeter 
cluster. The cluster energy $E$ is the sum of the cell energies. 
The cluster time $T$ and position $\vec{R}$ 
are energy weighted averages. Energy and time resolutions are $\sigma_E/E = 
5.7\%/\sqrt{E\ {\rm(GeV)}}$ and  $\sigma_t = 57\ {\rm ps}/\sqrt{E\ {\rm(GeV)}}
\oplus50\ {\rm ps}$, respectively.
The KLOE trigger \cite{TRG} uses calorimeter and chamber 
information. For this analysis only the calorimeter signals are relevant.
Two energy deposits with $E>50$ MeV for the barrel and  $E>150$ MeV for the 
endcaps are required.

Prompt photons are identified as neutral particles
with  $\beta=1$ originated at the 
interaction point  requiring 
$|T-R/c|<{\rm min}(5\sigma_T, 2\ {\rm ns})$, 
where $T$ is the photon flight time and $R$ the path length;
$\sigma_T$ includes also the contribution of the bunch length jitter.
The photon detection efficiency is $\sim 90\%$ for $E_\gamma$=20~MeV, and 
reaches 100\% above 70~MeV.
The sample selected by the timing requirement contains a $<1.8\%$ 
contamination due to accidental clusters from machine background.


Two amplitudes contribute to \phippg: \f\to$S$\gam, $S$\to\po\po\ ($S\gamma$)
and \f\to$\rho^0$\po, $\rho^0$\to\po\gam\ ($\rho\pi$) where $S$ is a scalar
meson.
The event selection criteria  of the \phippg\ decays (\ppg)
have been designed to give similar efficiencies 
for both processes. The first step,
requiring five prompt photons with $E_\gamma\!\ge$7 MeV and 
$\theta\ge\theta_{\rm min}=23^{\circ}$, reduces the sample
to 124,575 events.
The background due to \f\to\ks\kl\  is removed
requiring that $E_{\rm tot}=\sum_5 E_{\gamma,i}$ and 
$\vec p_{\rm tot}=\sum_5 \vec p_{\gamma,i}$ satisfy $E_{\rm tot}\!>$800 MeV 
and $|\vec p_{\rm tot}|\!<$200 MeV/c.
We are left with 15,825 events.
Other reactions which give rise to background are: 
\epm\to$\omega$\po\to\po\po\gam\
(\wpi), \f\to\et\po\gam\to5\gam\ (\et$\pi$\gam)
and \f\to\et\gam\to3\po\gam\ ($\eta\gamma$) with 2 undetected photons.

A kinematic fit (Fit1) requiring overall energy and momentum conservation
improves the energy resolution to $3\%$.
Photons are assigned to \pio's by minimizing a test $\chi^2$-function
for both the \ppg\ and \wpi\ cases. 
For the \wpi\ case we also require $M_{\pi\gamma}$ to be consistent with 
$M_{\omega}$.
The correct combination is found by this procedure 89\%, 96\% of the time
for the \ppg, \wpi\ case respectively.
Good agreement is found with the Monte Carlo simulation, MC, for the 
distributions of the $\chi^2$ and of the invariant masses. A second fit 
(Fit2) requires the masses of \gam\gam\ pairs to equal $M_{\pi}$.

\begin{figure}[!t]
\begin{center}
\includegraphics[width=0.45\textwidth]{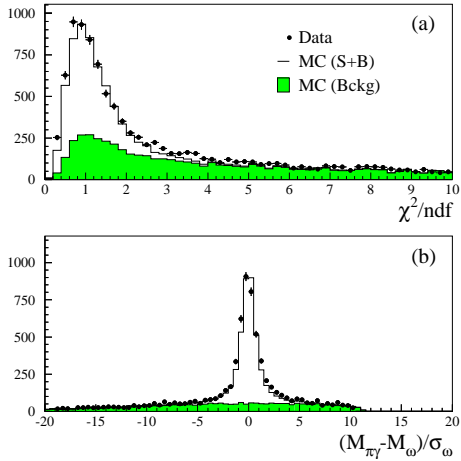}
\end{center}
\caption{Data--MC comparison for $\wpi$ events: (a) $\chi^2$/ndf and
(b) $\Delta M_{\pi\gamma}/\sigma_{\omega}$.}
\label{Fig:wpn}
\end{figure}

\begin{figure}[!t]
\begin{center}
\includegraphics[width=0.5\textwidth]{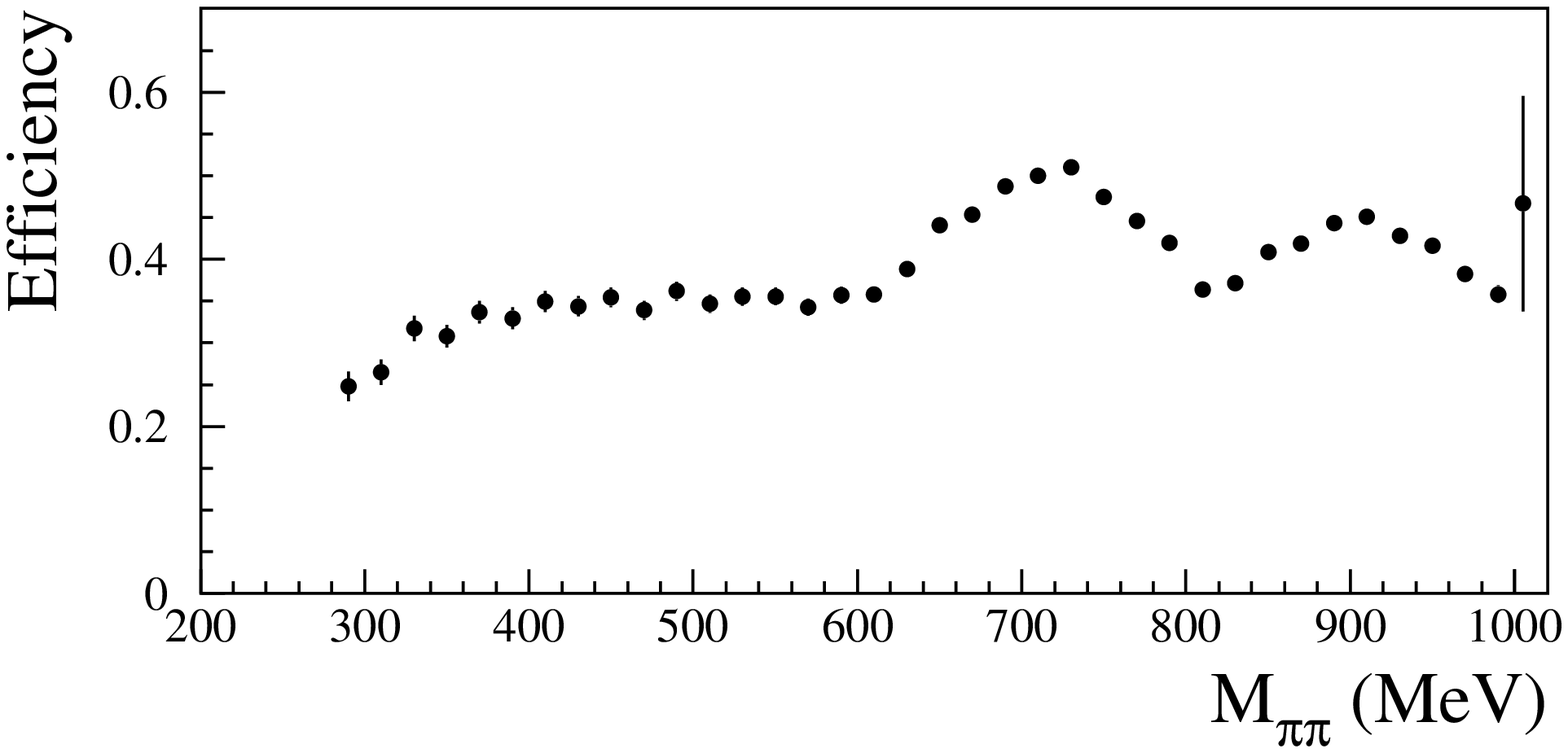}
\end{center}
\caption{Efficiency vs \pio\pio\ invariant mass for 
\phippg\ events.}
\label{Fig:Eff_vs_Mpp}
\end{figure}

\begin{figure}[!t]
\begin{center}
\includegraphics[width=0.6\textwidth]{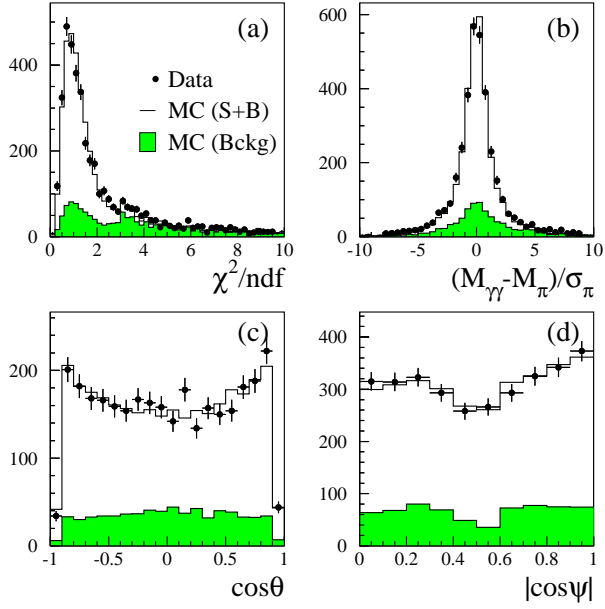}
\end{center}
\caption{Data--MC comparison for \phippg\ events after \wpi\ rejection: 
(a) $\chi^2$/ndf; (b) $(M_{\gam\gam}-M_{\pi})/\sigma_\pi$ with 
$\chi^2/{\rm ndf}\le 3$; (c, d) angular distributions with all analysis 
cuts applied. 
$\theta$ is the polar angle of the radiative photon, $\psi$ is the angle 
between the radiative photon and \pio\ in the \pio\pio\ rest frame.}
\label{Fig:f0n}
\end{figure}

\begin{figure}[!t]
\begin{center}
\includegraphics[width=0.6\textwidth]{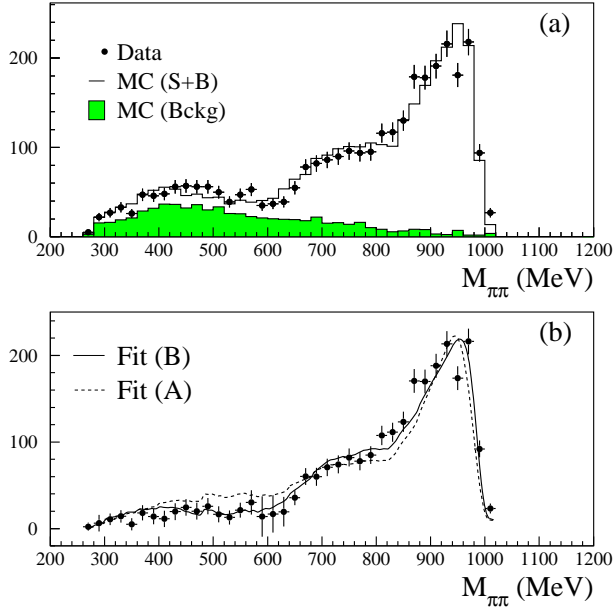}
\end{center}
\caption{Observed spectrum of \pio\pio\ invariant mass before (a) and after (b)
  background subtraction.}
\label{Fig:RawMpp}
\end{figure}

The \wpn\ background is reduced rejecting the events satisfying 
$\chi^2/{\rm ndf}\le 3$ and $\Delta M_{\pi\gamma}= |M_{\pi\gamma}-M_{\omega}|
\le 3\,\sigma_{\omega}$ using Fit2 in the \wpi\ hypothesis.
Data and MC are in good agreement (Fig.~\ref{Fig:wpn}.a-b). 
The \phippg\ events must then satisfy $\chi^2/{\rm ndf}\le 3$ for 
Fit2 in the \ppg\ hypothesis. We also require $\Delta M_{\gamma\gamma}= 
|M_{\gamma\gamma}-M_{\pi}| \le 5\,\sigma_{\pi}$ using the photon momenta 
of Fit1. 
The efficiency for the identification of the signal 
is evaluated applying the whole analysis chain 
to a sample of simulated \f\to$S$\gam, $S$\to\po\po\ events 
with a \pio\pio\ mass ($m$) spectrum consistent with the data. 
We use the symbol \mpp\ to denote the reconstructed value of $m$.
The selection efficiency as a function of  \mpp\ is shown in
Fig.~~\ref{Fig:Eff_vs_Mpp}. The average over
the whole mass spectrum is $\epsilon_{\pi\pi\gamma} = 41.6\%$.
A similar efficiency function is obtained for
the process $\phi\to\rho^0\pio$ with $\rho^0\to\pio\gamma$.
%
Fig.~\ref{Fig:f0n} shows various distributions for the 
3102 events surviving the selection  together
with MC predictions. 
The angular distributions prove that $S\gamma$ is the 
dominant process. 
The rejection factors and the expected
number of events for the background processes
are listed in Tab.~\ref{Tab:Bckg} \cite{KLOE_ao,KLOE_eg,KLOE_fo}.
\begin{table}[!t]
\caption{Background channels for \phippg.}
\label{Tab:Bckg}
\newcommand{\m}{\hphantom{$0$}}
\renewcommand{\tabcolsep}{0.6pc} 
\renewcommand{\arraystretch}{1.2} 
\begin{tabular}{@{}lcc} \hline
Process          &  Rejection Factor    & Expected events  \\ \hline
$\wpn$           &  8.7                 & $339\pm 24$      \\
$\phi\to\eta\pio\gamma\to\gamma\gamma\pio\gamma$ 
                 &  4.0                 & $166\pm 16$      \\
$\phi\to\eta\gamma\to\pio\pio\pio\gamma$     
                 &  $5.9\times 10^{3}$  & $159\pm 12$      \\ \hline
\end{tabular}
\end{table}
After subtracting the background 2438$\pm$61 \phippg\ events remain. 
Their \mpp\ spectrum is shown in Fig.~\ref{Fig:RawMpp}.

The \phippg\ branching ratio, \br, is obtained normalizing the number 
of events after background subtraction, $N-B$, to the $\phi$ cross 
section, $\sigma(\phi)$, to the selection efficiency and to \lint:

\begin{equation}
\br(\phippg) = 
  \frac{N-B}{\epsilon_{\pi\pi\gamma}}\:\frac{1}{\sigma(\phi)\,\lint}
\end{equation}
The luminosity is measured using large angle Bhabha scattering events.
The measurement of  $\sigma(\phi)$ is obtained from the 
$\phi\to\eta\gamma\to\gamma\gamma\gamma$ decay in the 
same sample \cite{KLOE_eg}. 
We obtain:

\begin{equation}
\br(\phippg) =
(1.08 \pm 0.03_{\rm stat} \pm 0.03_{\rm syst} \pm 0.04_{\rm norm})\x
10^{-4} .
\label{Eq:BRave}
\end{equation}
The contributions to the uncertainties are listed in Tab.~\ref{Tab:Errors}.
Details can be found in Ref.~\cite{KLOE_fo}.

\begin{table}[!t]
\caption{Uncertainties on BR(\phippg).}
\label{Tab:Errors}
\newcommand{\m}{\hphantom{$-$}}
\newcommand{\cc}[1]{\multicolumn{1}{c}{#1}}
\renewcommand{\tabcolsep}{2.pc} 
\renewcommand{\arraystretch}{1.2} 
\begin{tabular}{@{}lc} \hline
Source                 &  Relative error  \\ \hline
Statistics             &   2.5\%          \\
Background             &   1.3\%          \\
Event counting         &   2.3\%          \\
Normalization          &   3.7\%          \\ \hline
Total                  &   5.2\%          \\ \hline
\end{tabular}
\end{table}


In order to disentagle the contributions of the various processes and to 
determine the normalized differential decay rate,
$\dif\br/\dif m = (1/\Gamma)  \dif\Gamma/\dif m$, we fit the data to a 
mass spectrum $f(m)$.
This spectrum is taken as the sum of $S\gamma$, $\rho\pi$ and 
interference term, $f(m)=f_{S\gamma}(m) + f_{\rho\pi}(m) + f_{\rm int}(m)$.
The scalar term is \cite{AchIvanch}:
\begin{equation}
 f_{S\gamma}(m) = \frac{2\,m^2}{\pi}\:
 \frac{\Gamma_{\phi S\gamma}\Gamma_{S\pi^{0}\pi^{0}} }{|D_S|^2}\:
 \frac{1}{\Gamma_\phi}.
\end{equation}
The \f\to $S$\gam\ process is estimated by means of a $K^+K^-$ loop 
for the \fo:
\begin{equation}
\Gamma_{\phi f\!_{0}\gamma} (m) =
\frac{g^2_{f\!_0K^+K^-}g^2_{\phi K^+K^-}}{12\pi}\:
\frac{|g(m)|^2}{M_{\phi}^2}\:
\left( \frac{M^2_{\phi}-m^2}{2M_{\phi}} \right) ,
\end{equation}
where $g_{\phi K^+K^-}$ and $g_{f\!_{0} K^+K^-}$ are the couplings
and $g(m)$ is the loop integral function.

A recent measurement \cite{E791_sigma} reports the existence of a 
scalar $\sigma$ with $M_\sigma = (478^{+24}_{-23}\pm 17)$ MeV and 
$\Gamma_\sigma = (324^{+42}_{-40}\pm 21)$ MeV.
If we include the contribution of this meson, its decay rate is given 
by \cite{Ankara}:
\begin{equation}
\Gamma_{\phi\sigma\gamma} (m) =
\frac{e^2\,g^2_{\phi \sigma\gamma}}{12\pi}\:\frac{1}{M_\phi^2}\:
\left( \frac{M^2_{\phi}-m^2}{2M_{\phi}} \right)^3.
\end{equation}
where $g_{\f\sig\gam}$ is a point-like $\phi\sigma\gamma$ coupling.

Finally, $\Gamma_{S\pi^{0}\pi^{0}}$ is related to 
$\Gamma_{S \pi^{+}\pi^{-}}$ by: 
\begin{equation}
\Gamma_{S\pi^{0}\pi^{0}} (m) = 
\frac{1}{2} \,\Gamma_{S \pi^{+}\pi^{-}} (m) =
\frac{g^2_{S \pi^+\pi^-}}{32\,\pi\,m}\,\sqrt{1-\frac{4M_{\pi}^2}{m^2}}\ .
\end{equation}
For the inverse propagator, $D_S$, we use the formula with finite width 
corrections \cite{AchIvanch} for the \fo\ and a Breit Wigner for the 
$\sigma$. The parametrization of Ref.~\cite{AchInt} has been used for the 
$\rho\pi$ and the interference term.

The observed mass spectrum $S_{\rm obs}(\mpp)$ is fit folding 
into the theoretical shape experimental efficiency and resolution 
after proper normalization for $\sigma(\phi)$ and \lint . 
Two different fits have been performed
varying $f_{S\gamma}(m)$: in Fit (A) only the \fo\ contribution 
is considered while in Fit (B) we also include
the contribution of the $\sigma$ meson.
The mass and width of the $\sigma$ were fixed to their central values. 
If the normalization of the $\rho\pi$ term is left free during 
fitting, its contribution and the associated interference terms turn 
out to be negligibly small.
When $\br(\phi\to\rho^0\pio\to\pio\pio\gamma)$ is fixed at $1.8\cdot 10^{-5}$
as in Ref.~\cite{AchInt}, the $\chi^2$/ndf increases by more than a factor 
of 2.
The fits without the $\rho\pi$ contribution are shown superimposed over 
the raw spectrum in Fig.~\ref{Fig:RawMpp}.b.

The result of the fits are listed in Tab.~\ref{Tab:FitRes}.
\begin{table}[t]
\caption{Fit results using \fo\ only, Fit (A), and
including the $\sigma$, Fit (B).}
\label{Tab:FitRes}
\newcommand{\m}{\hphantom{$-$}}
\newcommand{\cc}[1]{\multicolumn{1}{c}{#1}}
\renewcommand{\tabcolsep}{0.8pc} 
\renewcommand{\arraystretch}{1.2} 
\begin{tabular}{@{}lcc} \hline
                            &  Fit (A)          &  Fit (B)           \\ \hline
$\chi^2/{\rm ndf}$          &  $109.53/34$      &  $43.15/33$        \\ 
$M\!\subfo$ (MeV)           &  $962\pm4$        &  $973  \pm 1$      \\
$g^2_{f\!_0 K^+K^-}/(4\pi)$ (GeV$^2$)       &  $1.29\pm0.14$    &  
  $2.79 \pm 0.12$   \\
$g^2_{f\!_0 K^+K^-}/g^2_{f\!_0\pi^+\pi^-}$  &  $3.22 \pm 0.29$  &  
  $4.00 \pm 0.14$   \\
$g_{\phi\sigma\gamma}$      &  ---              &  $0.060\pm 0.008$  \\ \hline
\end{tabular}
\end{table}
In Fit (A) we use as free parameters $M\!\subfo$, $g^2_{f\!_0 K^+K^-}$
and the ratio $g^2_{f\!_0 K^+K^-}/g^2_{f\!_0\pi^+\pi^-}$.
The fit gives a large $\chi^2$/ndf; integrating the theoretical spectrum 
a value $\br(\phifog\to\pio\pio\gamma) =
(1.11\pm 0.06_{\rm stat+syst})\x10^{-4}$ is obtained.

A much better agreement with data is given by Fit (B), where we add 
as a free parameter also the coupling $g_{\phi\sigma\gamma}$.
The negative interference between the \fo\ and $\sigma$ amplitudes results
in the observed decrease of the $\pio\pio\gamma$ yield below 700 MeV. 
Integrating over the theoretical $\sigma$ and \fo\ curves we obtain 
$\br(\phi\to\sigma\gamma\to\pio\pio\gamma) =
(0.28\pm0.04_{\rm stat+syst})\x10^{-4}$
and $\br(\phifog\to\pio\pio\gamma) = 
(1.49 \pm 0.07_{\rm stat+syst})\x10^{-4}$.
Multiplying the latter \br\ by a factor of 3 to account for 
$\fo\to\pip\pim$ decay, the $\br(\phifog)$ is determined to be:
\begin{equation}
\br(\phifog) = (4.47 \pm 0.21_{\rm stat+syst}) \cdot 10^{-4}.
\end{equation}

The values of the coupling constants from Fit (B) are in agreement with 
those reported by the SND and CMD-2 experiments \cite{SND_fo,CMD_fo}.
The coupling constants differ from the WA102 result on
\fo\ production in central $pp$ collisions 
($g^2_{f\!_0 K^{+}K^{-}}/g^2_{f\!_0\pi^{+}\pi^{-}} = g_K / 1.33\,g_{\pi} =
1.63 \pm 0.46$) \cite{WA102}
and from those obtained when the \fo\ is produced in 
$D_s^+\to\pip\pim\pip$ decays \cite{E791_fo}, where $g_K$ 
is consistent with zero.

In order to allow a detailed comparison with other experiments and
theoretical models, we have unfolded $S_{\rm obs}(\mpp)$.
For each reconstructed mass bin, the ratio between the theoretical and
the smeared function, $SF(\mpp)$, is calculated.
The $\dif\br/\dif m$ is then given by:
\begin{equation}
\frac{\dif\br}{\dif m} = 
  \frac{S_{\rm obs}(\mpp)}{SF(\mpp)}\:\frac{1}{\lint\x\sigma(\phi)\x\Delta\mpp}
\end{equation}

\begin{figure}[!t]
\begin{center}
\includegraphics[width=0.6\textwidth]{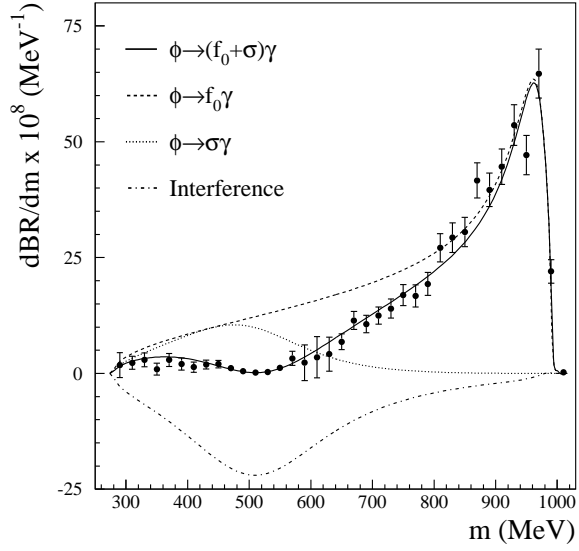}
\end{center}
\caption{$\dif\br/\dif m$ as a function of $m$. Fit (B) is shown as a solid 
         line; individual contributions are also shown.}
\label{Fig:dBRdMpp}
\end{figure}

The value of $\dif \br/\dif m$ as a function of $m$ is given in 
Tab.~\ref{Tab:dBRdMpp} and shown in Fig.~\ref{Fig:dBRdMpp}.
Integrating over the whole mass range we obtain:
\begin{equation}
\br(\phippg) =
(1.09 \pm 0.03_{\rm stat} \pm 0.03_{\rm syst} \pm 0.04_{\rm norm})\x
10^{-4} .
\end{equation}

which well compares with the result obtained correcting for 
the average selection efficiency (Eq.~\ref{Eq:BRave}).
If we limit the integration to the \fo\ dominated region, above 700~MeV,
we get:
$$\eqalign{
\br&(\phippg;\: m >700\ {\rm MeV})=\cr
&\kern3cm(0.96\pm0.02_{\rm stat}\pm0.02_{\rm syst}\pm0.04_{\rm norm})\x
10^{-4}.\cr}$$
which is in agreement with our previous measurement in the same mass range 
\cite{LP01}.

\begin{table}[!t]
\caption{Differential BR for \phippg. $m$ is expressed in MeV while 
$\dif{\rm BR}/\dif m$ is in units of $10^8\,{\rm MeV}^{-1}$.
The errors listed are the total uncertainties.}
\label{Tab:dBRdMpp}
\newcommand{\s}{\hphantom{1}}
\newcommand{\cc}[1]{\multicolumn{1}{c}{#1}}
\renewcommand{\tabcolsep}{1.0pc} 
\renewcommand{\arraystretch}{1.0} 
\vglue2mm
\begin{tabular}{@{}cc|cc} \hline
$m$        &   $\frac{d{\rm BR}}{dm}$   &   $m$  &
        $\frac{d{\rm BR}}{dm}$  \\ \hline
\s290      &    $2.0\pm 2.9$   &   \s670      &    $11.2\pm 1.9$     \\
\s310      &    $2.2\pm 1.4$   &   \s690      &    $11.0\pm 1.9$     \\
\s330      &    $3.0\pm 1.5$   &   \s710      &    $12.5\pm 1.9$     \\
\s350      &    $0.9\pm 1.3$   &   \s730      &    $14.0\pm 2.0$     \\
\s370      &    $2.9\pm 1.4$   &   \s750      &    $17.3\pm 2.3$     \\
\s390      &    $2.2\pm 1.3$   &   \s770      &    $17.0\pm 2.4$     \\
\s410      &    $1.4\pm 1.1$   &   \s790      &    $19.4\pm 2.5$     \\
\s430      &    $1.8\pm 1.0$   &   \s810      &    $27.4\pm 3.1$     \\
\s450      &    $1.9\pm 0.8$   &   \s830      &    $29.2\pm 3.2$     \\
\s470      &    $1.1\pm 0.5$   &   \s850      &    $30.6\pm 3.2$     \\
\s490      &    $0.5\pm 0.2$   &   \s870      &    $41.7\pm 3.8$     \\
\s510      &    $0.2\pm 0.1$   &   \s890      &    $39.6\pm 3.6$     \\
\s530      &    $0.3\pm 0.2$   &   \s910      &    $44.6\pm 3.8$     \\
\s550      &    $1.3\pm 0.5$   &   \s930      &    $53.6\pm 4.4$     \\
\s570      &    $3.3\pm 1.5$   &   \s950      &    $47.2\pm 4.3$     \\
\s590      &    $2.1\pm 3.6$   &   \s970      &    $64.7\pm 5.3$     \\
\s610      &    $3.7\pm 4.7$   &   \s990      &    $22.0\pm 2.5$     \\
\s630      &    $4.2\pm 3.7$   &    1010      &    $\s0.2\pm 0.1$    \\
\s650      &    $7.0\pm 1.7$   &              &                      \\ \hline
\end{tabular}
\end{table}

In a separate paper \cite{KLOE_ao}, we present a measurement of 
$\br(\phi\to a_0\gamma$), together with a discussion of the implications 
of \fo\ and $a_{0}$ results.

\section*{Acknowledgements}
%
We thank the DA$\Phi$NE team for their efforts in maintaining low
background running conditions and their collaboration during all
data-taking. We also thank F.~Fortugno for his efforts in ensuring
good operations of the KLOE computing facilities. 
We thank R.~Escribano for discussing with us
the existing theoretical framework.
This work was supported in part by DOE grant DE-FG-02-97ER41027; by 
EURODAPHNE, contract FMRX-CT98-0169; by the German Federal Ministry of 
Education and Research (BMBF) contract 06-KA-957; 
by Graduiertenkolleg `H.E. Phys.and Part. Astrophys.' of 
Deutsche Forschungsgemeinschaft, Contract No. GK 742;
by INTAS, contracts 96-624, 99-37; and by TARI, contract HPRI-CT-1999-00088.
%

%

\begin{thebibliography}{9}
\bibitem{Obs_fo}
  M.N.\,Achasov \etal, Phys.\ Lett.\ B 440 (1998) 442.
\bibitem{SND_fo}
  M.N.\,Achasov \etal, Phys.\ Lett.\ B 485 (2000) 349.
\bibitem{CMD_fo}
  R.R.\,Akhmetshin \etal, Phys.\ Lett.\ B 462 (1999) 380.
\bibitem{CIK}
  F.E.\,Close, N.\,Isgur and S.\,Kumano, Nucl.\ Phys.\ B 389 (1993) 513.
\bibitem{BrownClose}
  N.\,Brown and  F.E.\,Close, ``Second \dafne\ Physics Handbook'',
  Ed.\ L.\,Maiani, G.\,Pancheri and N.\,Paver, Vol.\ 2 (1995) 649.
\bibitem{Tornqvist}
  N.A.\,T\"{o}rnqvist, Phys.\ Rev.\ Lett.\ 49 (1982) 624.
\bibitem{Jaffe}
  R.L.\,Jaffe, Phys.\ Rev.\ D 15 (1997) 267.
\bibitem{Molecule}
  J.\,Weinstein and N.\,Isgur, Phys.\ Rev.\ Lett.\ 48 (1982) 659.
\bibitem{KLOE}
  KLOE Collaboration, LNF-92/019 (IR) (1992) and LNF-93/002 (IR) (1993).
\bibitem{DAFNE}
  S.\,Guiducci \etal, Proc.~of the 2001 Particle Accelerator Conference 
  (Chicago, Illinois, USA), P.\,Lucas S.\,Webber Eds., 2001, 353.
\bibitem{DCH}
  KLOE Collaboration, M.\,Adinolfi \etal,
  LNF Preprint LNF-01/016 (IR) (2001), accepted by Nucl.\ Inst.\ and Meth.
\bibitem{EMC}
  KLOE Collaboration, M.\,Adinolfi \etal,
  Nucl.\ Inst.\ and Meth.\ A 482 (2002) 363.
\bibitem{TRG}
  KLOE Collaboration, M.\,Adinolfi \etal,
  LNF Preprint LNF-02/002 (P) (2002), submitted to Nucl.\ Inst.\ and Meth.
\bibitem{KLOE_ao}
  KLOE Collaboration, A.\,Aloisio \etal, hep-ex/0204012 (2002),
  submitted to  Phys.\ Lett.\ B.
\bibitem{KLOE_eg}
  S.\,Giovannella and S.\,Miscetti, KLOE Note 177 (2002).
\bibitem{KLOE_fo}
  S.\,Giovannella and S.\,Miscetti, KLOE Note 178 (2002).
\bibitem{AchIvanch}
  N.N.\,Achasov and V.N.\,Ivanchenko, Nucl.\ Phys.\ B 315 (1989) 465.
\bibitem{E791_sigma}
  E.M.\,Aitala \etal, Phys.\ Rev.\ Lett.\ 86 (2001) 770.
\bibitem{Ankara}
  A.\,Gokalp and O.\,Yilmaz, Phys.\ Rev.\ D 64 (2001) 053017
  and private communication.
\bibitem{AchInt}
  N.N.\,Achasov and V.V.\,Gubin, Phys.\ Rev.\ D 63 (2001) 094007.
\bibitem{WA102} 
  D.\,Barberis \etal, Phys.\ Lett.\ B 462 (1999) 462;
  F.E.\,Close A.\,Kirk, Phys.\ Lett.\ B 515 (2001) 13.
\bibitem{E791_fo}
  E.M.\,Aitala \etal, Phys.\ Rev.\ Lett.\ 86 (2001) 765.
\bibitem{LP01}
  KLOE Collaboration, A.\,Aloisio \etal, hep-ex/0107024 (2001).
%
\end{thebibliography}
\end{document}